\documentclass[pra,letterpaper,twocolumn]{revtex4} 
\usepackage{amsbsy,amssymb,amsmath,bm,bbold} 
\usepackage{graphicx,color,epsfig,rotate} 
\usepackage{fancyhdr} 
\usepackage{epstopdf}
\usepackage{float}
\usepackage[colorlinks=true,linkcolor=blue,citecolor=blue,urlcolor=blue]{hyperref}

\def\bbbc{{\mathchoice {\setbox0=\hbox{$\displaystyle\rm C$}\hbox{\hbox 
to0pt{\kern0.4\wd0\vrule height0.9\ht0\hss}\box0}} 
{\setbox0=\hbox{$\textstyle\rm C$}\hbox{\hbox 
to0pt{\kern0.4\wd0\vrule height0.9\ht0\hss}\box0}} 
{\setbox0=\hbox{$\scriptstyle\rm C$}\hbox{\hbox 
to0pt{\kern0.4\wd0\vrule height0.9\ht0\hss}\box0}} 
{\setbox0=\hbox{$\scriptscriptstyle\rm C$}\hbox{\hbox 
to0pt{\kern0.4\wd0\vrule height0.9\ht0\hss}\box0}}}}

\begin{document} 
\title{Realization of Third Order Exceptional Singularities in a Three level non-Hermitian System: Towards Cascaded State Conversion} 
\author{Sayan Bhattacherjee, Arnab Laha,}
\thanks{Equal contributions}
\affiliation{Department of Physics, Indian Institute of Technology Jodhpur, Rajasthan-342037, India}
\author{Somnath Ghosh}
\email{somiit@rediffmail.com}
\affiliation{Department of Physics, Indian Institute of Technology Jodhpur, Rajasthan-342037, India}

\begin{abstract} 
	The appearance of topological singularities, namely \textit{exceptional points} (EPs) is an intriguing feature of parameter-dependent open quantum or wave systems. EPs are the special type of non-Hermitian degeneracies where two (or more) eigenstates of the underlying system coalesce. In this paper, we present a three level non-Hermitian Hamiltonian which hosts three interacting eigenstates. The matrix elements are optimized in such a way that the intermediate eigenstate interacts with both the other states and the underlying system hosts at least two different second order EPs. The impact of quasi-static parameter variation along a cyclic contour around the embedded EPs on the dynamics of interacting eigenvalues is well investigated in the context of cascaded state conversion. Such dynamics of the eigenvalues shows a clear signature of the third order EP with a combined effect of both the second order EPs. Moreover, we examine the accumulation of phases around the identified EPs and study the hallmark of phase exchange during cascaded state conversions accompanied by the parametric encirclement of the third order EP.  
\end{abstract} 
 
 
\maketitle %

\section{Introduction}

Instead of familiar Hermitian quantum systems, realistic open systems have always a great physical insight as they interact with the environment. These open quantum/ quantum inspired wave systems with metastable resonance states are well described by non-Hermitian formalism in quantum mechanics. The parametric dependence of interaction phenomena between the complex resonances of a non-Hermitian system~\cite{Kato,Sannino} is now a central focus in many domains of physics, especially in optics and photonics. In this context, a key-point is the phenomenon of \textit{avoided resonance crossing} (ARC) between the resonances in complex eigenvalue-plane with parameter dependent crossing/ anti-crossing of their frequencies and widths (i.e., essentially the real and imaginary parts of the complex eigenvalues respectively)~\cite{Kato,Sannino}. A specific interesting feature of an ARC can be realized by associating a particular second-order branch-point singularity in the parameter spectrum. At this singular point, the interacting eigenvalues approach a special type of degeneracy; which is characteristically far different from genuine Hermitian degeneracies. Such a specific spectral singularity is coined as \textit{exceptional point} (EP) by T. Kato~\cite{Kato}. The EPs appear as topological defects usually in open quantum systems that depend on at least two real (or a complex) parameters. At an EP, the interacting eigenvalues (two or more) and the corresponding eigenstates of the underlying Hamiltonian simultaneously coalesce and after coalescence, the eigenstates lose their identities and pickup a huge magnitude.  

Over the past decade, the fascinating features of EPs in open quantum systems have taken tremendous attention in various domains of physics. Intensive theoretical efforts have been put forward to observe the signature of EPs in atomic~\cite{Cartarius1} as well as molecular spectrum~\cite{Lefebvre}, microwave billiards~\cite{Dietz}, system with cold atoms~\cite{Xu}, coupled asymmetric dimers~\cite{Jin}, etc. The topological structure of an EP has been experimentally demonstrated via analogous study between the Schr\"odinger and the Helmholtz equations in microwave cavity with an explicit observation of a chiral state~\cite{Dembowski1,Dembowski2}. Specifically, in the optical domain the unconventional physical effects of EPs have been explored theoretically as well as experimentally in various open photonic systems {\it viz.} optical waveguides~\cite{Doppler,Chen G,Goldzak,Zhang,Midya,Ghosh}, microcavities~\cite{Laha2,Wiersig2,Chen W,Phang,Yi,Lee SB}, laser systems~\cite{Liertzer,Hodaei2,Hodaei3}, photonic crystals~\cite{Ding1,Bykov}, etc. The specific connection of a special category of EP with broken $\mathcal{PT}$-symmetry~\cite{Ganainy,Wang} is now well established when various open systems are operated under the $\mathcal{PT}$-symmetric restrictions~\cite{Xu,Jin,Goldzak,Zhang,Phang,Hodaei3,Ding1}.

A deficient hallmark of the appearance of EPs is the violation of adiabatic theorem for parametric encirclement around it along with a closed contour; where EPs lead crucial modifications in dynamics of the corresponding coupled resonances. Following a quasi-static cyclic movement of control parameters around an EP, the corresponding decaying eigenstates transfer all its population to their coupled counterparts. If the EP is inside the parametric loop then the coupled eigenvalues are adiabatically transformed into each other along with the simultaneous transformation of corresponding eigenvectors accompanying an additional phase shift~\cite{Dembowski1}, while all other eigenstates regain their initial states at end of the loop. Such effect of encirclement around a second order EP have theoretically well established~\cite{Uzdin,Milburn} in the contextual phenomena of asymmetric mode conversion~\cite{Ghosh}, selective flip-of-states~\cite{Laha2}, cross-polarization mode switching~\cite{Midya,Bykov}, etc and also verified experimentally~\cite{Dembowski1,Dembowski2,Doppler}. The chiral behavior of the eigenfunctions~\cite{Harney} around an EP has also been discussed in the context of quantum phase transitions (including geometric phase as well as dynamical phases) during state switching~\cite{Muller,Eleuch}; where during encirclement, accumulated geometric phase differs from the Barry phase~\cite{Mailybaev}. Such dramatic behaviors of the coupled states near EPs play a key role in various contemporary technological applications and generic phenomena like dark state lasing~\cite{Hodaei2}, extreme enhancement in sensing~\cite{Wiersig2,Chen W}, etc.

While, most of the reported works highlight two levels coalescence at a second order branch point as the second order EP; recently there are evolving interests towards more than two levels coalescence and recognition of higher order EPs~\cite{Heiss4,Demange,Eleuch1,Ryu}. In this context, three levels coalescence with realization of a third order EP has richer physical insight and technological impact in comparison with two levels coalescence at a second order EP. To distinguish the order of EPs, here we abbreviate a second order EP as EP2 and a third order EP of as EP3.  Considering such a system having more than two interacting states, one can identify multiple EP2s~\cite{Eleuch1,Ryu,Shallem,Lee SY}. It has been generally observed that, for independent controlling of $m$ interacting states $(m^2+m-2)/2$ parameters are needed~\cite{Heiss4}; where with proper manipulation of required parameters $m(m-1)/2$ EP2s can be encountered. Now more importantly, with combined effects of $(N-1)$ EP2s, an analogous effect of an $N$-fold EP may be realized~\cite{Muller}. Thus for $m=3$ (i.e. for a system with three mutually coupled states) three additional parameters along with two previously chosen parameters (i.e. total five parameters) are needed to control the interactions towards encounter an EP3. In this situation, with judicious manipulation of coupling parameters, \textit{one state must have to couple with the rest of two states and analytically connected with them via two square root branch points (i.e. two EP2s)}. Here, with combined effect of two EP2s, an analogous EP3 can be realized where three levels are analytically connected by a cube root branch point; while this phenomena is far different from a traditional three-fold degeneracy usually occur in Hermitian systems. To avoid this misconception, it is always preferred to understand an EP3 as a coalescence of two EP2; where three eigenvalues coalesce~\cite{Heiss4}. Such coalescence of EP2s can be achieved with proper variation of three additional parameters.

Lately, attempts have been made to encounter, understood and explore the physical properties of an EP3 theoretically~\cite{Heiss4,Demange,Eleuch1,Ryu,Shallem,Lee SY,Heiss2}; and studied in different open systems like optical microcavity~\cite{Lee SY,Hodaei1}, waveguide~\cite{Heiss5}, photonic crystal~\cite{Lin2}, Bose-Einstein gases system~\cite{Gutohrlein}, Bose-Hubbard system~\cite{Graefe}, atomic system~\cite{Menke}, etc. The coalescence of multiple EP2s has experimentally demonstrated in an acoustic cavity~\cite{Ding2}. Technologically, EP3 has been proposed to be utilized enhance sensitivity extremely in comparison with an EP2 in the context of EP based microcavity sensors~\cite{Hodaei1}. Moreover, various proposals have been reported towards the effect of parametric encirclement around an EP3 which may be realized if the connected EP2s are simultaneously enclosed in a single closed contour in respective parameter plane~\cite{Gutohrlein}. \textit{In this context, systematic analysis, encounter and direct observation of higher order state-conversion is yet to be explored; where both the hallmarks of two EP2s and an EP3 should be clearly manifested}. If realized such study should open up a new platform for a whole new range of photonic devices including integrated mode-converters, circulators, mode-multiplexers, etc.

In this paper, we realize a three-state open system with consideration of a three level non-Hermitian matrix. Here, the passive system having three decaying eigenstates is subjected to a perturbation. The perturbation matrix consists of required five effective coupling parameters. Judiciously manipulating the control parameters, one specific state is deliberately chosen to interact with rest of the states and analytically connect with them via two EP2s. Encircling these two EP2s in respective parameter plane, we study the dynamics of the coupled eigenvalues in the context of higher order state conversion in complex eigenvalue plane in the vicinity of an {\it analogous} EP3. Here to the best of our knowledge, we propose a mathematical model for prototype designing to encounter directly the third order EP with associated hallmark features for the first time.  \textit{Exploring subsequent state conversions, we propose an exclusive flip-of-states phenomena for the first time, exploiting EP3 as a third order branch point for eigenvalues}. Recently, EPs have attracted considerable attention due to their fascinating relation with quantum phase switching~\cite{Lee SY}. In this context, we also calculate the accumulation of phase picked up during this parametric encirclement and identify phase switching between respective coupled state around EP3. From these investigations, we explore a specific signature of EP3 in the context of higher order state conversion. With precise parametric optimizations, proposed scheme may be implemented in various realistic quantum inspired or wave based systems for device level applications.

\section{Three level non-Hermitian system: Analytical Model}

To study the fascinating topological characteristics of an EP3, we study the situation of a three level coalescence by considering a three state open system. Accordingly, we construct a simple realistic $3\times3$ non-Hermitian Hamiltonian matrix $\mathcal{H}$; where a passive Hamiltonian $H_0$ having three independent decaying eigenstates is subjected to a parameter dependent complex perturbation $H_p$. The complete Hamiltonian $\mathcal{H}$ having the form $H_0+\lambda H_p$ is represented as follows. 
\begin{equation}
\mathcal{H}=\left(\begin{array}{ccc}\widetilde{\varepsilon}_1 & 0 & 0 \\0 & \widetilde{\varepsilon}_2 & 0 \\0 & 0 & \widetilde{\varepsilon}_3\end {array}\right)+\lambda\left(\begin{array}{ccc}0 & \delta-\gamma  & 0 \\ \kappa & 0 & \gamma \\ 0 & \delta-\kappa & 0\end {array}\right).
\label{equation_H}
\end{equation}
Here, $\lambda\, (=\lambda_R+i\lambda_I)$ is a complex tunable parameter. In the passive Hamiltonian $H_0$, $\widetilde{\varepsilon}_j\,(j=1,2,3)$ are the three complex passive eigenvalues; where we consider $\widetilde{\varepsilon}_j=\varepsilon_j+i\tau_j\,(\tau_j<<\varepsilon_j)$. Here, $\varepsilon_j$ represent three real passive eigenvalues with corresponding small decay rates $\tau_j$. In the perturbation matrix, $\gamma$ and $\kappa$ are two real coupling terms; which are connected through a tunable real parameter $\delta$. Thus including complex $\lambda$, we have total five parameters to evoke the interactions between the eigenvalues of $\mathcal{H}$, denoting as $E_j;\,j=1,2,3$. The elements of the perturbation matrix are chosen and optimized in such a way that for a fixed $\gamma$ and $\kappa$, $\delta$ is able to control the coupling phenomena between $E_1$ and $E_2$ as well as $E_2$ and $E_3$ independently, over an complex independent parameter $\lambda$. Thus, $E_2$ should be coupled with both $E_1$ and $E_3$ at two different ($\delta,\lambda$)-regions. Here, we don't consider the interaction between $E_1$ and $E_3$ deliberately.

Now,  $E_j$ (three eigenvalues of the Hamiltonian $\mathcal{H}$) are obtained from the roots of the following cubic secular equation, given as
\begin{equation}
E^3+\alpha_1E^2+\alpha_2E+\alpha_3=0;
\label{equation_E}
\end{equation}
where,
\begin{subequations}
	\label{equation_alpha}
	\begin{align}
	\alpha_1&=-(\widetilde{\varepsilon}_1+\widetilde{\varepsilon}_2+\widetilde{\varepsilon}_3),\\
	\begin{split}
	\alpha_2 
	&=\widetilde{\varepsilon}_1\widetilde{\varepsilon}_2+\widetilde{\varepsilon}_2\widetilde{\varepsilon}_3+\widetilde{\varepsilon}_3\widetilde{\varepsilon}_1\\
	&\qquad-\lambda^2\{\gamma(\delta-\kappa)+\kappa(\delta-\gamma)\},
	\end{split}\\
	\alpha_3&=-\widetilde{\varepsilon}_1\widetilde{\varepsilon}_2\widetilde{\varepsilon}_3+\lambda^2\{\gamma(\delta-\kappa)\widetilde{\varepsilon}_1+\kappa(\delta-\gamma)\widetilde{\varepsilon}_3\}.
	\end{align}
\end{subequations}
Using Cardano's method~\cite{Korn}, the roots of the Eq.~\ref{equation_E}, i.e. the eigenvalues of the Hamiltonian $\mathcal{H}$ can be written as
\begin{subequations}
	\label{equation_Eig}	
	\begin{align}
	E_1&=\omega\epsilon_++\bar{\omega}\epsilon_--\eta,
	\label{equation_Eig1}\\
	E_2&=\epsilon_++\epsilon_--\eta,
	\label{equation_Eig2}\\
	E_3&=\bar{\omega}\epsilon_++\omega\epsilon_--\eta;
	\label{equation_Eig3}
	\end{align}
\end{subequations}
with
\begin{equation}
	\epsilon_{\pm}=(m\pm \sqrt{m^2+n^3})^{1/3}\quad\textnormal{and}\quad\eta=\alpha_1/3.
	\label{equation_epsilon-eta} 
\end{equation}
Here, $\omega$ is the cube root of unity where $\omega^3=1$; $\bar\omega$ is the complex conjugate of $\omega$. Now, $m$ and $n$ can be written in terms of $\alpha_j\,;j=1,2,3$ (given in the Eqs.~\ref{equation_alpha}) as
\begin{subequations}
	\begin{align}
	m&=-\alpha_1^2/27+\alpha_1\alpha_2/6-\alpha_3/6,\\
	n&=-\alpha_1^2/9+\alpha_2/3.
	\end{align}
\end{subequations}
Now, one can identify two different EP2s associated with the coalescence between $E_1$ and $E_2$ as well as between $E_2$ and $E_3$ at two different ($\delta,\lambda$)-regions. Such situations take place if the following conditions are fulfilled.
\begin{equation}
\epsilon_+=\epsilon_-\quad\textnormal{and}\quad\omega\epsilon_+=\epsilon_-\,\,\textnormal{or}\,\,\bar{\omega}\epsilon_+=\epsilon_-.
\label{equation_EP}
\end{equation}
The equalities in Eqs.~\ref{equation_EP} satisfy if the square root part of $\epsilon_{\pm}$ (as given in Eq.~\ref{equation_epsilon-eta}) is vanished. Thus the cube root nature of $\epsilon_{\pm}$ under the conditions for occurrence of two different EP2s implies that the three eigenvalues as given by the Eqs.~\ref{equation_Eig} are analytically connected by an analogous cube root branch point, i.e. an EP3; which can be realized with coalescence of two identified EP2s.

In the following sections, based on our proposed Hamiltonian $\mathcal{H}$ as given by Eq.~\ref{equation_H}, we execute a numerical study on realizing the situation of three levels coalescence via an EP3 with a combined effect of two EP2s and an exclusive application towards cascaded flip-of-states phenomena around an EP3. During optimization, we choose $\varepsilon_1=0.76$, $\varepsilon_2=0.65$ and $\varepsilon_3=0.3$; where the corresponding decay rates $\tau_1=0.005$, $\tau_2=0.0025$ and $\tau_3=0.0002$. To make the mathematical model inclusive and feasible as a prototype, we have considered the elements of the passive matrix to be complex (as $\widetilde{\varepsilon}_j=\varepsilon_j+i\tau_j$) having the extremely small imaginary parts ($\tau_j$) in comparison to the real parts ($\varepsilon_j$). In the perturbation part of the $\mathcal{H}$, we choose only the real values of $\gamma$ and $\kappa$ as $\gamma=0.95$ and $\kappa=0.3$ respectively.
 
\section{Exceptional Points in the framework of the proposed non-Hermitian Hamiltonian}         

\subsection{Avoided resonance crossing and encountering multiple EP2s}

\begin{figure}[t]
	\centering
	\includegraphics[width=8.8cm]{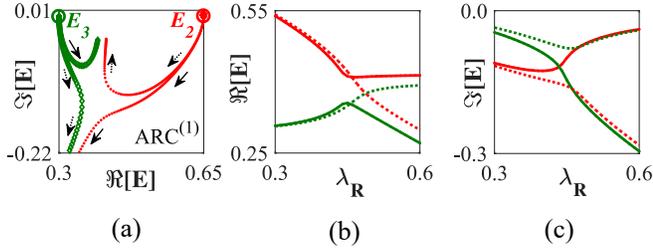}
	\caption{(Color online) \textbf{(a)} Trajectories of $E_2$ and $E_3$, depicted by evolution of red dotted and green diamond markers, exhibiting ARCs in complex $E$-plane. The direction of evolutions are shown by dotted arrows for $\delta=0.21$ and solid arrows for $\delta=0.23$. Red and green big circular markers represent the initial positions of $E_2$ and $E_3$ respectively. \textbf{(b)} Corresponding crossing/ anti-crossing in $\Re[E]$ and \textbf{(c)} simultaneous anti-crossing/ crossing in $\Im[E]$ with increase in $\lambda_R$. Such variations of $\Re[E]$ and $\Im[E]$ of $E_2$ and $E_3$ with respect to $\lambda_R$ are shown by dotted red and green lines for $\delta=0.21$; whereas the same for $\delta=0.23$ are depicted by solid red and green lines respectively.}
	\label{figure_arc1}
\end{figure}

\begin{figure}[t]
	\centering
	\includegraphics[width=8.8cm]{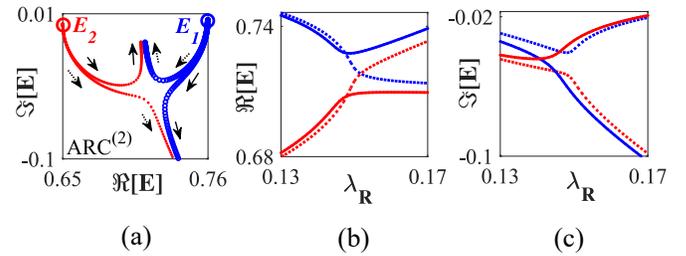}
	\caption{(Color online) \textbf{(a)} ARCs between $E_1$ and $E_2$, depicted by evolution of blue circular and red dotted markers, along the directions shown by dotted arrows for $\delta=1.26$ and solid arrows for $\delta=1.29$. Blue and red big circular markers represent the initial positions of $E_1$ and $E_2$ respectively. \textbf{(b)} Corresponding crossing/ anti-crossing in $\Re[E]$ and \textbf{(c)} simultaneous anti-crossing/ crossing in $\Im[E]$ with increase in $\lambda_R$. Such variations of $\Re[E]$ and $\Im[E]$ of $E_1$ and $E_2$ with respect to $\lambda_R$ are shown by dotted blue and red green lines for $\delta=1.26$; whereas the same for $\delta=1.29$ are depicted by blue and red solid lines respectively.}
	\label{figure_arc2}
\end{figure}

\begin{figure}[b]
	\centering
	\includegraphics[width=8.8cm]{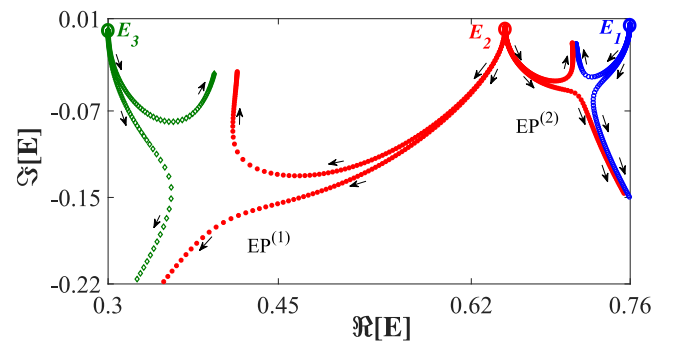}
	\caption{(Color online) Three-state-ARC associated with all the coupled states; where $E_2$ interacts simultaneously with $E_1$ and $E_3$. Trajectories of $E_j$ with $j = 1,2,3$ are shown by evolutions of blue circular, red dotted and green diamond shaped markers. Black crosses represent the influenced regions due to presence of two EP2s.}
	\label{figure_arcboth}
\end{figure}

Considering the optimized parametric values, here we study the mutual interactions between $E_j\,(j=1,2,3)$ through the phenomena of special ARCs with suitable adjustment of a real parameter $\delta$ over a complex parameter $\lambda\,(=\lambda_R+i\lambda_I)$. Now, to investigate such mutual interactions, we study the dynamics of $E_1$, $E_2$ and $E_3$ in the complex eigenvalue-plane ($E$-plane) with an quasi-static complex variation of $\lambda$ in a specified range for different $\delta$-values. Here, $\lambda_R$ varies from 0 to 0.6 with simultaneous and almost similar variation of $\lambda_I$ in the same range. Associated phenomena of ARCs are depicted in the Figs.~\ref{figure_arc1}--\ref{figure_arcboth}. Interestingly, for gradual increase in $\lambda$ and $\delta$, it is observed that $E_2$ interacts with $E_3$ at higher $\lambda$ and lower $\delta$ for which $E_1$ is unaffected; whereas for lower $\lambda$ and higher $\delta$, $E_ 2$ interacts with $E_1$, keeping $E_3$ as an observer. In this context, if we tune only one parameter between $\lambda_R$ and $\lambda_I$, fixing other, then $E_2$ is only able to interact with either $E_1$ or $E_3$ based on the choices of other parameters. Hence owing to simultaneous tuning in both $\lambda_R$ and $\lambda_I$, a controlled interaction phenomena is possible. Thus we deliberately choose the same range for both $\lambda_R$ and $\lambda_I$ in which they may vary independently. This is the crucial restriction we impose on the model.
        
Now in the Fig.~\ref{figure_arc1}, the interactions between $E_2$ and $E_3$ are depicted via a special ARC phenomena for two distinct values of $\delta$ over a continuous slow variation of $\lambda$. We judiciously choose such two $\delta$-values for which $E_2$ is going to interact with $E_3$ keeping $E_1$ unaffected. For $\delta=0.21$, they exhibit ARC in complex $E$-plane as can be seen in Fig.~\ref{figure_arc1}(a); where evolutions of $E_2$ and $E_3$ are shown by red dot and green diamond markers respectively; and directed by dotted arrows. Here $\Re[E]$ experiences a crossing and simultaneously $\Im[E]$ undergoes an anti-crossing with increase in $\lambda$ as depicted in Fig.~\ref{figure_arc1}(b) and Fig.~\ref{figure_arc1}(c) respectively; where the variations are shown only with respect to $\lambda_R$ (one may obtain the similar crossing/anti-crossing behaviour with respect to $\lambda_I$ also).
Here, the variations of $\Re[E]$ and $\Im[E]$ are marked by dotted red and green lines for $E_2$ and $E_3$ respectively. Now, while we slightly increase $\delta$ and fix at 0.23, $E_2$ and $E_3$ display a different kind of ARC in Fig.~\ref{figure_arc1}(a) where the  trajectories exchange their identities as directed by solid arrows. In this case, $\Re[E]$ undergoes an anti-crossing with simultaneous crossing in $\Im[E]$ as shown in Fig.~\ref{figure_arc1}(b) and Fig.~\ref{figure_arc1}(c) respectively. Here the variations of $\Re[E]$ and $\Im[E]$ are denoted by red and green solid lines for $E_2$ and $E_3$ respectively. Thus the behaviors of ARCs between $E_2$ and $E_3$ for two different $\delta$-values are topologically dissimilar. There must be a sudden transition between $\delta=0.21$ and $\delta=0.23$ where the coupled states coalesce while passing through a square root branch point singularity. It is evident that this singular point must be an EP2 which will appear in ($\delta,\lambda_R$)-plane where the coupled states are analytically connected~\cite{Ghosh,Laha2}. The approximate $\delta$-coordinate of the identified EP2 can be obtained with proper scanning $\delta$-values closer to the special point; whereas the approximate $\lambda_R$-coordinate can be found out by taking an average between $\lambda_R$-coordinates of the crossing-points of $\Re[E]$ and $\Im[E]$ from Fig.~\ref{figure_arc1}(b) and Fig.~\ref{figure_arc1}(c) respectively. Here the approximate location of the EP2 is identified at $\sim$(0.22, 0.45) in ($\delta,\lambda_R$)-plane; which is denoted as EP2$^{(1)}$ in the following text.

In the Fig.~\ref{figure_arc2}, we study the similar ARC phenomena for two specified higher $\delta$-values (in comparison with the previously chosen values) where $E_2$ is going to couple with $E_1$ keeping $E_3$ as an observer.
In the complex $E$-plane the evolutions of $E_1$ and $E_2$ are marked by blue circular and red dotted markers. Now for a fixed $\delta=1.26$, the coupled states encounter an ARC along the directions shown by dotted arrows in Fig.~\ref{figure_arc2}(a) with crossing in $\Re[E]$ and simultaneous anti-crossing in $\Im[E]$ with increase in $\lambda_R$ as displayed by dotted blue and red lines (for $E_1$ and $E_2$ respectively) in Fig.~\ref{figure_arc2}(b) and Fig.~\ref{figure_arc2}(c) respectively. For slight higher value of $\delta$ as 1.29, the trajectories of $E_1$ and $E_2$ exchange their directions shown by solid arrows in Fig.~\ref{figure_arc2}(a) and exhibit ARC; where $\Re[E]$ undergoes an anti-crossing and simultaneously $\Im[E]$ experiences a crossing as can be seen in Fig.~\ref{figure_arc2}(b) and Fig.~\ref{figure_arc2}(c) respectively by solid blue and red lines (for $E_1$ and $E_2$ respectively). Similarly the heterogeneous behaviour of ARCs between $E_1$ and $E_2$ for two different chosen $\delta$ as 1.26 and 1.29 evidently discloses the presence of another EP2 at $\sim(\delta=1.275,\lambda_R=0.15)$; which is denoted as EP2$^{(2)}$.

\begin{figure*}[htpb]
	\centering
	\includegraphics[width=16.5cm]{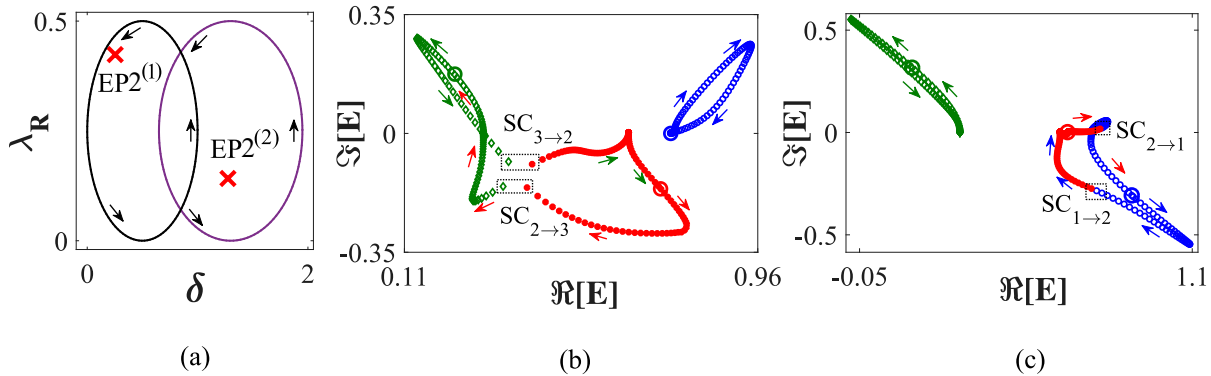}
	\caption{(Color online) \textbf{(a)} Individual encirclements around EP2$^{(1)}$ and EP2$^{(2)}$ in ($\delta$,$\lambda_R$)-plane as shown by black and violet contours respectively. The red crosses represent the approximate positions of two EP2s. For the black contour we choose a center at (0.5,0.25) and the characteristics parameters $a=1=b$, whereas for the violet contour the center is chosen at (1.25,0.25) with $a=0.5$ and $b=1$.  \textbf{(b)} Dynamics of $E_j\,(j=1,2,3)$ in complex $E$-plane corresponding to the black contour in (a). \textbf{(c)} Similar trajectories of $E_j\,(j=1,2,3)$ corresponding to the violet contour in (a). In both (b) and (c), the dynamics of $E_1$, $E_2$ and $E_3$ represented by the evolutions of blue circular, red dotted and green diamond shaped markers; where big circles with respective colors represent their initial positions at the beginning of the encirclement. The state-conversion phenomena in the $E$-plane are clearly visible inside the dotted rectangles, where SC$_{i\rightarrow j}$ ($i,j=1,2,3;\,i\ne j$) means the {\it state conversion} from i$^{th}$ state to j$^{th}$ state. The arrows indicate the direction of progressions where the colors of the arrows represent the journey of the respective states from their initial to final positions.}
	\label{figure_oneEP}
\end{figure*}

\subsection{Three-state-ARC}

The overall behaviour of mutually interacting eigenvalues and whole ARC phenomena is presented in Fig.~\ref{figure_arcboth}, where $E_2$ interacts simultaneously with $E_1$ and $E_3$. Here, we behold two situations with respect to choice of $\delta$ over the specified $\lambda$-span, where among three interacting states any two are coupled keeping the third as an observer; while interacting states exhibit special ARCs at two different ($\delta,\lambda_R$)-regions. Thus we have identified two EP2s in ($\delta,\lambda_R$)-plane. We may now explore if exists, the signature of the presence/ physical insight of an cube root branch point, i.e., an EP3; where three states are analytically connected via the combined effect of two EP2s~\cite{Eleuch1}. In Fig.~\ref{figure_arcboth}, the approximate influenced regions in complex $E$-plane due to presence of two identified EP2s are shown by black cross type markers.

\section{Numerical findings: Effect of parametric encirclements around EP2s}

In this section, to establish the exact second order singular behaviours of the identified EP2s, we study the unconventional physical properties via adiabatic parametric encirclement around them towards cascaded state conversion. Hence, to study such effects of parametric encirclement, we consider a closed loop following the coupled equations given as
\begin{subequations}
	\label{equation_circle} 
	\begin{align}
	\delta(\theta)&=x_{0}\left[1+a\,\cos(\theta)\right],\\
	\lambda_R(\theta)&=y_{0}\left[1+b\,\sin(\theta)\right].
	\end{align}
\end{subequations}
where ($x_0,y_0$) is the center. $a$ and $b$ are two characteristics parameters which control the variations of $\delta$ and $\lambda_R$ with a tunable angle $\theta$. Here $\{a,b\}\in(0,1]$ and $\theta\in[0,2\pi]$. Here $\lambda_I$ is also tuned simultaneously in the almost same variation-range of $\lambda_R$. As per requirements, one may judiciously choose an arbitrary center ($x_0,y_0$) and two characteristics parameters $a$ and $b$ to encounter single or both/ multiple EP2s in ($\delta$,$\lambda_R$)-plane and scan the enclosed area.

\subsection{Encircling EP2s individually}

\begin{figure*}[htbp]
	\centering
	\includegraphics[width=14cm]{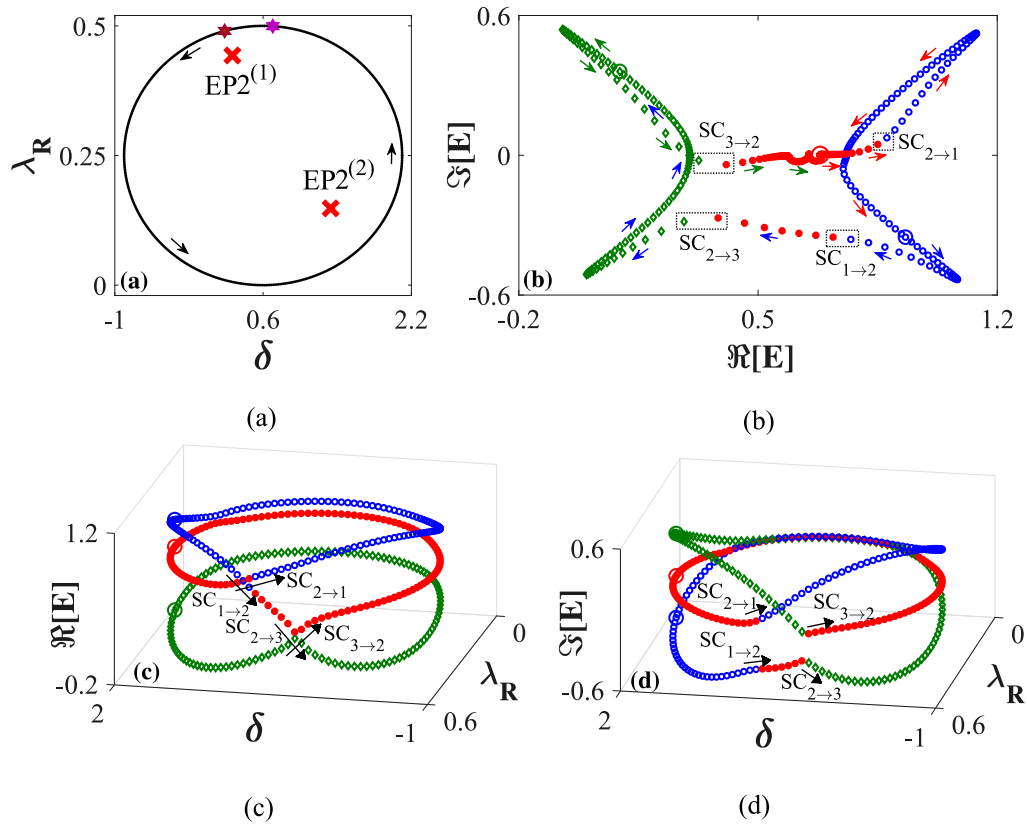}
	\caption{(Color online) \textbf{(a)} Parametric encirclement process in ($\delta$,$\lambda_R$)-plane enclosing both EP2$^{(1)}$ and EP2$^{(2)}$. Here the circular loop is chosen having center at (0.6,0.25) with $a=2.5$ and $b=1$. Here magenta and brown stars indicate two specific locations of $\theta$ as $0.46\pi$ and $0.582\pi$, where conversions take place. \textbf{(b)} Associated trajectories of $E_j\,(j=1,2,3)$ in complex $E$-plane. Magenta star in (a) correspond SC$_{1\rightarrow 2}$ and SC$_{2\rightarrow 1}$; and brown star in (a) correspond SC$_{2\rightarrow 3}$ and SC$_{3\rightarrow 2}$. Individual distributions of \textbf{(c)} $\Re[E]$ and \textbf{(d)} $\Im[E]$  with respect to the circular variations of $\delta$ and $\lambda_R$ along the black contour in (a). Here the black arrows indicate the evaluation directions followed by subsequent state conversions. Other used notations, colors and markers carry the similar meaning as described in Fig.~\ref{figure_oneEP}.}       
	\label{figure_twoEP}
\end{figure*}
Here, we encircle the identified EP2s individually in the ($\delta$,$\lambda_R$)-plane to check the dynamics of mutually coupled states ($E_j,\,j=1,2,3$) in complex $E$-plane. At first, we choose a closed contour in parameter plane in such a way that it encloses the only EP2$^{(1)}$, keeping EP2$^{(2)}$ out of the described loop; which is shown by black curve in Fig.~\ref{figure_oneEP}(a). Associated dynamics of the coupled states $E_2$ and $E_3$; and also the observer state $E_1$ are displayed in Fig.~\ref{figure_oneEP}(b). The trajectories of $E_j\,(j=1,2,3)$ are shown by blue circular, red dotted and green diamond shaped markers respectively; where the big circles of respective colors represent their initial positions in complex $E$-plane. Each point on the trajectories of the every states in $E$-plane corresponds to the analogous each point on enclosed contour in ($\delta$,$\lambda_R$)-plane. As can be seen in Fig.~\ref{figure_oneEP}(b), following the anti-clockwise encirclement along the black circular loop around EP2$^{(1)}$ (as shown in Fig.~\ref{figure_oneEP}(a)), all three states start moving from their initial positions; where $E_2$ and $E_3$ are moving towards each other and suddenly make a conversion between them, while $E_1$ moves unaffectedly. In Fig.~\ref{figure_oneEP}(b), the conversions between the individual states are clearly displayed inside the dotted rectangles where SC$_{i\rightarrow j}$ ($i,j=1,2,3;\,i\ne j$) means the {\it state conversion} from i$^{th}$ state to j$^{th}$ state. After one round encirclement, $E_2$ is going to the starting position of $E_3$ and $E_3$ takes the initial location of $E_2$; and make a complete loop in complex $E$-plane, whereas $E_1$ makes an individual isolated loop and return to its own initial position. Now it is observed that after next encirclement $E_2$ and $E_3$ regain their initial positions following same trajectories and similar conversions exhibiting EP2$^{(1)}$ as seconder order branch point for eigenvalues, while $E_1$ makes an complete loop along the same individual path again.

Secondly, we consider a different contour in ($\delta$,$\lambda_R$)-plane which enclose only EP2$^{(2)}$ as shown by violet circular trajectory in Fig.~\ref{figure_oneEP}(a) and study the dynamics of all the three states in Fig.~\ref{figure_oneEP}(c), where the used notations, colors, markers carry the same meaning as we described in Fig.~\ref{figure_oneEP}(b). Here interestingly we observe that for one round complete encirclement in anti-clockwise direction around EP2$^{(2)}$ in parameter plane, there are conversions between $E_1$ and $E_2$; and after conversion they exchange their initial positions making a complete loop in $E$-plane, while $E_3$ remains unaffected and makes an individual loop. After next encirclement $E_1$ and $E_2$ retrieve their initial positions. $E_3$ remains isolated in this case.

Thus from the Fig.~\ref{figure_oneEP}, it is evident that when only EP2$^{(1)}$ is enclosed by the parametric encirclement then only $E_2$ and $E_3$ show the mutual conversion between them keeping $E_1$ as an observer in complex $E$-plane; because instead of $E_1$ only $E_2$ and $E_3$ are analytically connected through an EP2$^{(1)}$.  Similar conversion phenomena between $E_1$ and $E_2$ happens in the $E$-plane keeping $E_3$ as an observer when only EP2$^{(2)}$ is rightly encircled in ($\delta$,$\lambda_R$)-plane. Hence it is established that both the EP2$^{(1)}$ and EP2$^{(2)}$ exhibit exact second order singular behavior even in presence of an nearby third state in a three-state system owing to the special coupling restrictions.

\subsection{Encircling multiple EP2s: State conversion around EP3}                    

Here, we implement the encircling scheme as given in Eq.~\ref{equation_circle} and enclose both EP2$^{(1)}$ and EP2$^{(2)}$ in same parametric loop. The arbitrary center ($x_0,y_0$) and the characteristics parameters $a$ and $b$ are chosen accordingly (as given in the caption of Fig.~\ref{figure_twoEP}). Following such encirclement process in ($\delta$,$\lambda_R$)-plane as shown by black contour in Fig.~\ref{figure_twoEP}(a), the dynamics of the mutually coupled eigenvalues $E_j\,(j=1,2,3)$ are displayed in Fig.~\ref{figure_twoEP}(b). Used notations, colors, markers carry the previously described specific meanings.

Now, tracking the quasi-static anticlockwise encirclement around both the EP2s in ($\delta$,$\lambda_R$)-plane through enough small steps, $E_j$ start moving from their initial locations (i.e. the locations for $\theta=0$) with increase in $\theta$. Now, when $\theta=0.46\pi$ (the location marked by magenta star on the black contour in Fig.~\ref{figure_twoEP}(a)) then for corresponding $\delta$ and $\lambda_R$ values, $E_1$ is converted to $E_2$ and also $E_2$ is converted to $E_1$, however, there is no conversion in $E_3$. The conversions between $E_1$ and $E_2$ are shown by the notations SC$_{1\rightarrow 2}$ and SC$_{2\rightarrow 1}$ respectively in Fig.~\ref{figure_twoEP}(b). Now for further increase in $\theta$ through very small steps, when we reach $\theta=0.582\pi$ (the location marked by brown star on the black contour in Fig.~\ref{figure_twoEP}(a)), $E_2$ is converted to $E_3$ and also vice-versa, i.e. $E_3$ is converted to $E_2$; the locations of which are shown by the notations SC$_{2\rightarrow 3}$ and SC$_{3\rightarrow 2}$ respectively in complex $E$-plane. Accomplishing these subsequent conversions, it is observed that at the end of the first round encirclement, $E_1$ goes to the location of $E_3$ through an additional conversion with $E_2$, while $E_3$ and $E_2$ directly flip in the locations of $E_2$ and $E_1$ respectively and make a complete loop in complex $E$-plane. Thus there are three successive state flipping between $E_1$, $E_2$ and $E_3$. In Fig.~\ref{figure_twoEP}(c) and Fig.~\ref{figure_twoEP}(d), we plot the individual dynamics of $\Re[E]$ and $\Im[E]$ respectively with respect to the circular variation of $\delta$ and $\lambda_R$ along the closed contour shown in Fig.~\ref{figure_twoEP}(a). The subsequent all conversions phenomena as shown in Fig.~\ref{figure_twoEP}(b) are also visible clearly near the black arrows in Fig.~\ref{figure_twoEP}(c) and Fig.~\ref{figure_twoEP}(d). Under the special coupling restrictions, the simultaneous presence of two EP2s inside the parametric loop, to a complete surprise as reveled a new physical effect. The analytical connections between the EP2s through the third state, forms the origin of the third order branch point (EP3).    

\begin{figure}[t]
	\centering
	\includegraphics[width=8.65cm]{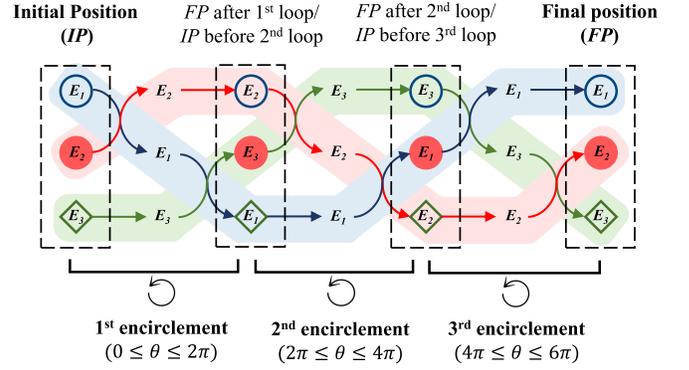}
	\caption{(Color online) Schematic representation of the flip-of-states phenomena in the vicinity of an EP3 for three successive parametric encirclements ($0\le\theta\le 6\pi$) around two EP2s along a closed contour. The initial locations of $E_1$, $E_2$ and $E_3$ are represented by blue hollow circle, red solid circle and green hollow diamond respectively. The arrows with the respective colors represent the journey of each state from $\theta=0$ to $\theta=6\pi$.}
	\label{figure_schematic}
\end{figure}

Exploiting the previously described state conversion scheme, in the Fig.~\ref{figure_schematic}, we represent an exclusive flip-of-states phenomena schematically for three successive encirclements around both the identified EP2s. At the beginning of the 1$^{st}$ encirclement, the initial locations of $E_1$, $E_2$ and $E_3$ are represented by blue hollow circle, red solid circle and green hollow diamond respectively. After completion of the 1$^{st}$ encirclement, they successively flip their locations as demonstrated numerically in Fig.~\ref{figure_twoEP} and schematically in Fig.~\ref{figure_schematic}. As $E_1$ and $E_2$ are analytically connected through EP2$^{(2)}$ they flip their locations directly after completion of the 1$^{st}$ encirclement. Similarly, there is direct flipping between $E_2$ and $E_3$ as they are connected via EP2$^{(1)}$.
However, in case of the permutation between the locations of $E_1$ and $E_3$, they transit through an additional conversion via $E_2$ due to the combined effect of both EP2$^{(1)}$ and EP2$^{(2)}$. Thus the presence of an EP3 is confirmed; where three states are analytically connected with realization of an cube root branch point. Now the final positions of all the states after 1$^{st}$ encirclement are carried forward as initial locations before the 2$^{nd}$ round encirclement along the same contour. Here interestingly, during the 2$^{nd}$ round, $E_2$ feels the effect of EP3 and permute its location with the position of $E_1$ (consider the location before beginning of the 2$^{nd}$ loop) through an additional conversion with $E_3$; whereas $E_1$ and $E_3$ are directly flipped to $E_3$ and $E_2$ respectively. In similar way, during 3$^{rd}$ round encirclement, $E_3$ experiences the effect of EP3 and permutes with the location of $E_2$ (location after end of the 2$^{nd}$ loop) via a conversion with $E_1$. Here, $E_2$ and $E_1$ are directly flipped to $E_1$ and $E_3$ respectively. Increasingly, it is evident that after completion of the 3$^{rd}$ round parametric encirclement, all the three mutually coupled states $E_1$, $E_2$ and $E_3$ retrieve their extreme initial positions, i.e., the locations at the beginning of the 1$^{st}$ encirclement. Thus, here an EP3 exhibits itself as the third order branch point for eigenvalues.

\begin{figure}[t]
	\centering
	\includegraphics[width=8.8cm]{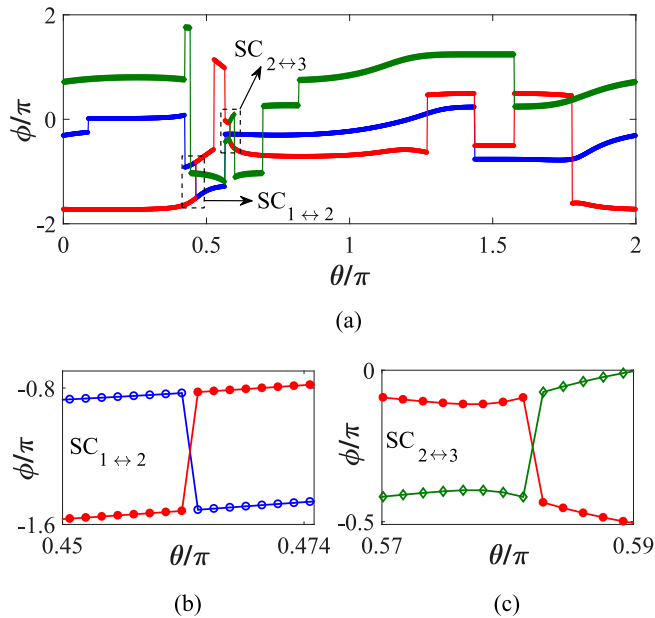}
	\caption{(Color online) \textbf{(a)} Phase variations $\phi_j\,(j=1,2,3)$ of three individual states. $\phi_1$, $\phi_2$ and $\phi_3$ are denoted by blue, red and green line with blue circular, red dotted and green diamond shaped markers. The phase switching near $\theta=0.46\pi$ and $\theta=0.582\pi$ are shown inside the first and second dotted rectangular boxes (from left side). \textbf{(b)} Zoomed view of the first dotted rectangle shown in (a) near $\theta=0.46\pi$, where we omit the variation of $\phi_3$ for clear visualization of the phase transition between $\phi_1$ and $\phi_2$. \textbf{(c)} Zoomed view of the second dotted rectangle shown in (a) near $\theta=0.582\pi$, where we omit the variation of $\phi_1$ for proper visibility of the transition between $\phi_2$ and $\phi_3$. More importantly, the notation SC$_{i\leftrightarrow j}$ correspond to the both SC$_{i\rightarrow j}$ and SC$_{j\rightarrow i}$ as described in Fig.~\ref{figure_twoEP}(b).}
	\label{figure_phase}
\end{figure}

Now, we study the phase variations of the mutually coupled states $E_j\,(j=1,2,3)$ for parametric encirclement process in ($\delta$,$\lambda_R$)-plane along the closed loop as shown in Fig.~\ref{figure_twoEP}(a), which enclose both the EP2$^{(1)}$ and EP2$^{(2)}$. Accordingly, we calculate the corresponding eigenfunctions say, $\psi_j\,(j=1,2,3)$ for each ($\delta$,$\lambda_R$) values on the black parametric contour and plot the associated accumulated phases (say, $\phi_j$) with respect to the cyclic angle ($\theta$) in Fig.~\ref{figure_phase}.
In the Fig.~\ref{figure_phase}(a) the variations of $\phi_1$, $\phi_2$ and $\phi_3$ are shown by blue, red and green lines with the specific markers of the respective style and colors as described in previous figures. Here, it is conspicuous that after one round encirclement around the EP2s, the phase of the system restored, i.e. either 0 or $2\pi$ (the accumulated $\phi_j$ are equal for $\theta=0$ and $\theta=2\pi$). Now we specifically focus on the phase switching phenomena during state conversions in complex $E$-plane followed by the described parametric encirclement. In Fig.~\ref{figure_twoEP}, we have shown that for $\theta=0.46\pi$, conversions take place between $E_1$ and $E_2$. Consequently, a clear phase switching is observed between $\phi_1$ and $\phi_2$ at the same $\theta$-value which is shown inside the first dotted rectangular box in Fig.~\ref{figure_phase}(a) and also zoomed in Fig.~\ref{figure_phase}(b). Here the notation SC$_{i\leftrightarrow j}$ correspond the two simultaneous state conversions as can be seen in Fig.~\ref{figure_twoEP}(a), i.e., from i$^{th}$ state to j$^{th}$ state and also the vice-versa. Similarly, for $\theta=0.582\pi$, there is phase transition between $\phi_2$ and $\phi_3$ as shown inside the second dotted rectangle in Fig.~\ref{figure_phase}(a) and zoomed in Fig.~\ref{figure_phase}(c); where in the $E$-plane conversions take place between $E_2$ and $E_3$. In the Fig.~\ref{figure_phase}(b), for the zoomed view we omit the variation of $\phi_3$ and in Fig.~\ref{figure_phase}(c)  we omit the variation of $\phi_1$ for clear visualizations. Thus it is evident that for one round parametric encirclement (i.e. for $0\le\theta\le 2\pi$), there are two phase switching of $\psi_2$ with respect to both $\psi_1$ and $\psi_3$ for two different $\theta$-values. This is the clear signature of the fact that $E_2$ is simultaneously connected with both $E_1$ and $E_3$ with two different EP2s; where all of them are analytically connected through an omnipresent EP3.

\section{Conclusion}

In summary, we investigate the effect of existence of multiple EP2s in a three state non-Hermitian Hamiltonian system, hosting three interacting eigenvalues; which is properly modeled with judicious choice of the matrix elements. The proposed Hamiltonian exhibits two special ARCs associated with three interacting states in complex eigenvalue-plane; where one state is analytically connected with the rest of two states via two EP2s in 2D parameter-plane. For such mutual coupling phenomena we identify the required total five different parameters. Here a real parameter ($\delta$) with specific tunability controls the simultaneous variations of two fixed coupling parameters ($\gamma$ and $\kappa$). Now, $\delta$ is able to control and manipulate two EP2s via special ARCs with another complex parameter $\lambda\,(=\lambda_R+i\lambda_I)$. Here $\lambda_R$ is exploited to locate two EP2s with $\delta$, and $\lambda_I$ connects two identified EP2s deliberately; which remarkably encounters the presence of a hidden EP3 in chosen parameter plane for the very first time. Now, to explore the state conversion phenomena in the vicinity of an EP3, we study the effect of parametric encirclement around both the identified EP2s via scanning the respective enclosed areas at a time, on the dynamics of the mutually coupled states. Hence exploiting such conversion schemes, we propose an exclusive flip-of-states phenomena which evidents successive switching among three corresponding states in complex eigenvalue-plane for the first time; establishing an EP3 as a third order branch point for eigenvalues. We also examine the accumulated phase variation of three coupled states and report the corresponding phase exchange during conversion of individual states. Our study numerically revels the specific relationship between coupling control parameters and the state conversion phenomena around an EP3 with simultaneous signature of two individual EP2s. We may design and optimize a prototype wave-based system to device and implement such third order special singular points. Hence state manipulation in such open systems will open up a new platform for integrated optical devices. Straight-forward measurement of accumulated phases related to states would also enable us to identify the existence of any such higher order EPs in the system during operation.   

\section*{Acknowledgments} 

SB acknowledges the financial support from MHRD. AL and SG acknowledges the financial support by the Department of Science and Technology (DST), Ministry of Science and Technology under INSPIRE Faculty Fellow Grant (IFA-12, PH-23).

\end{document}